\documentclass[twocolumn,prl,showpacs,preprintnumbers,amsmath,
amssymb]{revtex4}

\usepackage[dvips]{graphicx}
\usepackage{dcolumn}
\usepackage{bm}
\begin{document}
%\preprint{APS/123-QED}

\title{Nature of magnetism in Ca$_3$Co$_2$O$_6$}

\author{Hua Wu, M. W. Haverkort, D. I. Khomskii, and L. H. Tjeng}
\address{ II. Physikalisches Institut, Universit{\"a}t zu K{\"o}ln,
  Z{\"u}lpicher Str. 77, D-50937 K{\"o}ln, Germany}

\date{\today}

\begin{abstract}
We find using LSDA+U band structure calculations that the novel
one-dimensional cobaltate Ca$_3$Co$_2$O$_6$ is not a
ferromagnetic half-metal but a Mott insulator. Both the octahedral
and the trigonal Co ions are formally trivalent, with the
octahedral being in the low-spin and the trigonal in the
high-spin state. The inclusion of the spin-orbit coupling leads
to the occupation of the minority-spin $d_{2}$ orbital for the
unusually coordinated trigonal Co, producing a giant orbital
moment (1.57 $\mu_{B}$). It also results in an anomalously large
magnetocrystalline anisotropy (of order 70 meV), elucidating why
the magnetism is highly Ising-like. The role of the oxygen holes,
carrying an induced magnetic moment of 0.13 $\mu_{B}$ per oxygen, 
for the exchange interactions is discussed.
\end{abstract}


\maketitle

The one-dimensional (1D) cobaltate Ca$_3$Co$_2$O$_6$
\cite{Fjellvag96} and isostructural $A_3MM'$O$_6$
($A$=alkaline-earth metal; $M$,$M'$=transition metals) compounds
\cite{Niitaka99} attract now considerable attention due to their
peculiar crystal structure and very unusual properties. A large
number of research activities have been triggered by the
discovery of the unconventional magnetic structure with strange
magnetization jumps
\cite{Aasland97,Kageyama97,Maignan00,Martinez01,Niitaka01,Flahaut04,Maignan04,Hardy04,Samp04,Sekimoto04},
reminiscent of quantum tunneling of magnetization in molecular
magnets, and also by the observation of the large thermoelectric
power and magnetoresistance \cite{Raquet02,Hardy03,Maignan03}.

From a theoretical point of view, the understanding of the
Ca$_3$Co$_2$O$_6$ system is far from satisfactory. The crystal
structure consists of triangular lattice of $c$-axis chains of
alternating CoO$_6$ octahedra and trigonal prisms sharing common
faces \cite{Fjellvag96,Aasland97}. Both the spin and valence
state of the Co ions in these two positions, i.e. Co$_{oct}$ and
Co$_{trig}$, respectively, are a matter of controversy. There were
several suggestions made
\cite{Fjellvag96,Niitaka99,Aasland97,Kageyama97,Maignan00,Martinez01,
Niitaka01,Flahaut04,Maignan04,Hardy04,Samp04,Sekimoto04,Raquet02,
Hardy03,Maignan03,Takubo05},
the most common ones being either the non-magnetic low-spin (LS,
S=0) Co$^{3+}_{oct}$ and magnetic high-spin (HS, S=2)
Co$^{3+}_{trig}$, or the LS (S=1/2) Co$^{4+}_{oct}$ and HS
(S=3/2) Co$^{2+}_{trig}$. Recent LSDA and GGA calculations do not
provide clarity: Whangbo \textit{et al.} \cite{Whangbo03} and
Eyert \textit{et al.} \cite{Eyert04} support the first
alternative, while Vidya \textit{et al.} \cite{Vidya03} the
second one. To add to the confusion, all those calculations
\cite{Whangbo03,Eyert04,Vidya03} predict Ca$_3 $Co$_2$O$_6$ to be
a metal, in strong disagreement with the experiments
\cite{Maignan00,Raquet02,Maignan03}. In fact, it has been even
claimed \cite{Vidya03} that this material would be the first 1D
oxide displaying ferromagnetic half-metallicity (FMHM).

Another problem is the nature of magnetism in this material. In
modeling the intriguing magnetic properties, it was always
concluded or assumed that the Co chains behave as an Ising system
with ferromagnetic (FM) intrachain exchange
\cite{Aasland97,Kageyama97,Maignan00,Martinez01,Niitaka01,Flahaut04,Maignan04,Hardy04}.
The nature of such behavior, however, is completely unclear: the
origin of Ising-like behavior was not discussed at all, and
attempts to explain the FM exchange using an ionic
superexchange model \cite{Fresard04} meet some problems.

In view of these controversies, we carried out a theoretical
study of the electronic structure and magnetic properties of
Ca$_3$Co$_2$O$_6$ in which we took more explicitly into account
the correlated motion of the electrons typical in transition metal
oxides. In this work we used the LSDA+U method \cite{Anisimov91},
including also the spin-orbit coupling (SOC). We settle theoretically
the issue on the valence (spin) state, and find that
Ca$_3$Co$_2$O$_6$ is a Mott or rather charge-transfer insulator
\cite{ZSA} already for moderate values of Hubbard $U$.

%%disproving
%%previous theoretical studies \cite{Whangbo03,Eyert04,Vidya03}.

The inclusion of the SOC leads to a series of surprising but
experimentally sound results: a giant orbital moment for the HS
Co$_{trig}$ ion consistent with the experimentally observed large
total magnetic moment, an anomalously large magnetocrystalline
anisotropy important to explain the Ising-like
properties, and an unusual orbital occupation of the Co$_{trig}$
ion requiring a completely new approach for modeling the
exchange interactions.

Our calculations were performed by using the full-potential
augmented plane waves plus local orbital method \cite{WIEN2k}. We
took the Ca$_3$Co$_2$O$_6$ structure data determined by
neutron diffraction measurements at 40 K \cite{Fjellvag96}. The
muffin-tin sphere radii are chosen to be 2.7, 2.0 and 1.6 Bohr
for Ca, Co and O atoms, respectively. The cut-off energy of 16
Ryd is used for plane wave expansion,
% of interstitial wave functions, 
and 5$\times$5$\times$5 {\bf k}-mesh for integrations
over the Brillouin zone. 
%both of which ensuring the sufficient numerical accuracy. 
The SOC is included by the
second-variational method with scalar relativistic wave functions
\cite{WIEN2k}. The easy magnetization direction is along the
$c$-axis chains.

\begin{figure}[h]
 \centering\includegraphics[width=8cm]{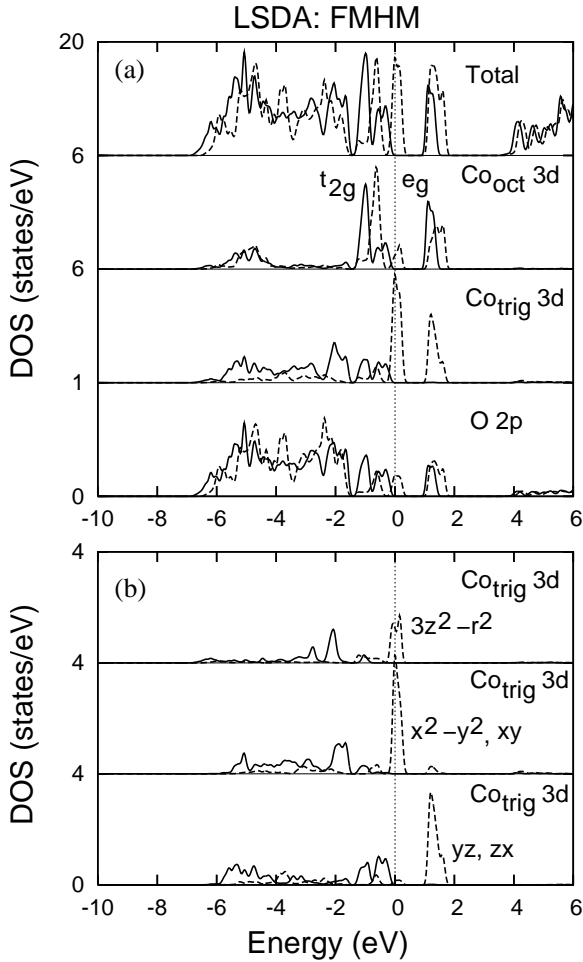}
 \caption{Density of states (DOS) of Ca$_3$Co$_2$O$_6$ in the
 ferromagnetic half-metallic (FMHM) state from LSDA. The Fermi level is set
 at zero energy. (a) shows (from upper to lower panels) the total DOS,
 the Co$_{oct}$ $3d$, Co$_{trig}$ $3d$, and O $2p$ partial DOS.
 (b) shows the Co$_{trig}$ $3d$ orbitally-resolved partial DOS.
 Solid (dashed) lines depict the spin-up (down) states.}
 \label{fig1}
\end{figure}

We plot in Fig. 1 the LSDA results, which give a FMHM solution,
in agreement with previous calculations
\cite{Whangbo03,Eyert04,Vidya03}. 
%We also find in our
%calculations that the Co$_{oct}$ and Co$_{trig}$ ions have both
%the 3+ valence, with the Co$_{oct}$ in the LS (S=0) and
%Co$_{trig}$ in the HS (S=2) state, confirming the results of
%Whangbo \textit{et al.} \cite{Whangbo03} and Eyert \textit{et
%al.} \cite{Eyert04}.  We certainly cannot support the
%Co$^{4+}_{oct}$/Co$^{2+}_{trig}$ scenario of Vidya \textit{et al.}
%\cite{Vidya03}, since we find that the Co$_{oct}$ ion in fact has
%even slightly more electrons ($\approx$0.3$e$) than the
%Co$_{trig}$. While it is clear that the LSDA scheme cannot
%reproduce the insulating nature of this compound, 
One should note that the Fermi level is located in a narrow Co$_{trig}$
$3d$ band with no more than 1 eV width, which is consistent with
the system being 1D. It is then also natural to expect that already 
modest electron correlation effects at the Co sites will be able
to turn this material into a Mott insulator.

\begin{figure}[h]
 \centering\includegraphics[width=8cm]{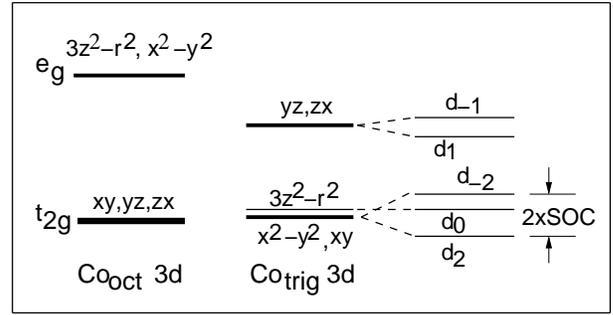}
 \caption{Local crystal field energy diagram for: (left) Co$_{oct}$ and
  (right) Co$_{trig}$ without and with spin-orbit coupling.}
 \label{fig2}
\end{figure}

Another important point to note from our LSDA results is that for
the Co$_{trig}$ ion, the crystal fields do not split up the $3d$
orbitals into the usual $e_{g}$ and $t_{2g}$ levels. Instead, it is
found that the $x^{2}-y^{2}$ is degenerate with the $xy$, and the
$yz$ with $zx$ orbital, as can be seen from Fig. 1(b). 
In the presence of the SOC, it is then better to use the
complex orbitals $d_0$, $d_{2}$/$d_{-2}$, and $d_{1}$/$d_{-1}$.
Significant from the LSDA is also that the narrow $d_0$ singlet
and $d_2$/$d_{-2}$ doublet bands are almost degenerate. Fig. 2
sketches the local crystal field energy diagram. This implies
that SOC, so far not included in all the LSDA calculations
\cite{Whangbo03,Eyert04,Vidya03}, could have a substantial effect
on the outcome for the predictions on the magnetic
properties. An interesting aspect from Fig. 1(b) is the
finding that the $d_1$/$d_{-1}$ band is split off from the $d_0$
and $d_2$/$d_{-2}$ bands by about 1 eV, i.e. much larger than the
SOC energy scale. Therefore, the SOC Hamiltonian can be simplified
into just $\zeta$$l_zs_z$ by neglecting the $l_{+}s_{-}$ and
$l_{-}s_{+}$ mixing terms. This has far reaching consequences for the
magnetocrystalline anisotropy as we will show below.

In Fig. 3 we show our LSDA+U+SOC results \cite{note}. We
find that Ca$_3$Co$_2$O$_6$ is an insulator
with ferromagnetic chains (FMI), i.e. not a metal, in
contrast to the LSDA results
\cite{Whangbo03,Eyert04,Vidya03}. We have used $U$=5 eV for the
Coulomb energy and $J_H$=0.9 eV for the Hund exchange parameter.
These values are taken from electron spectroscopy measurements
\cite{Saitoh95} and Hartree-Fock calculations \cite{Tanaka94}. To
investigate to what extent the insulating solution is robust
against the particular choice of $U$, we have calculated and
plotted in Fig. 4 (top panel) the band gap as a function of $U$.
One can observe that a value as small as 2.5 eV is already
sufficient to open up a gap. This is consistent with the system
being 1D resulting in a small width of the Co $3d$ band.The presence 
of this gap does not practically depend on the inclusion of SOC and 
is due to the Hubbard term $U$. We
therefore can conclude that Ca$_3$Co$_2$O$_6$ is a 
correlated insulator \cite{ZSA} for any realistic value of
$U$, i.e. 4---6 eV for Co$^{3+}$.

\begin{figure}[h]
 \centering\includegraphics[width=8cm]{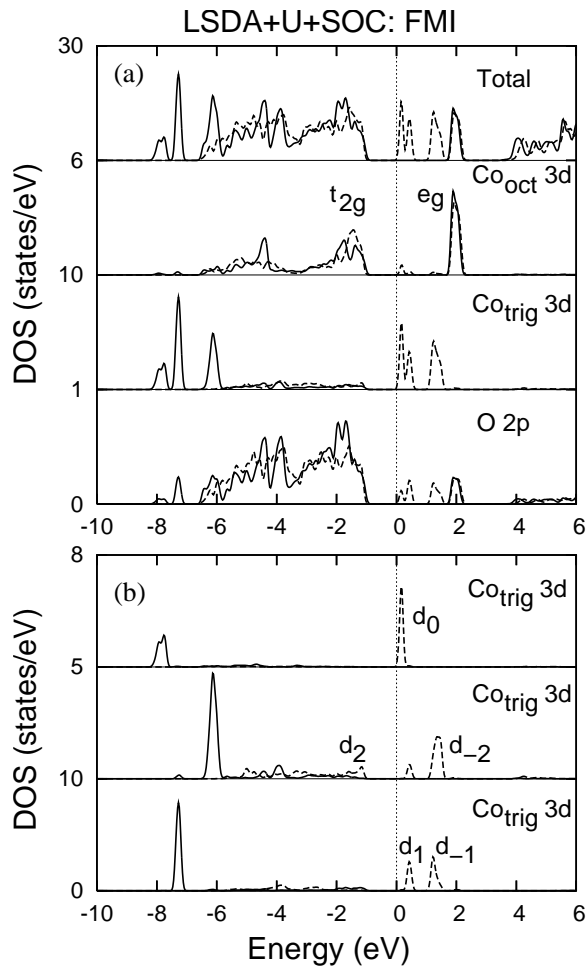}
 \caption{Density of states (DOS) of Ca$_3$Co$_2$O$_6$ in the
 ferromagnetic insulating (FMI) state from LSDA+U+SOC.}
 \label{fig3}
\end{figure}

\begin{figure}[h]
 \centering\includegraphics[width=8cm]{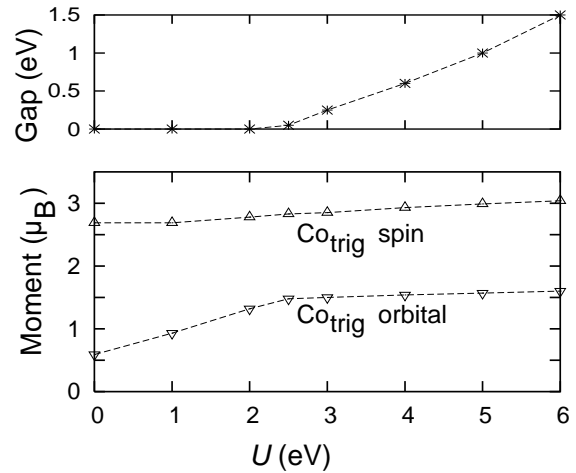}
 \caption{$U$ dependence of the insulating gap (top panel), as well as
  of the spin and orbital contributions to the Co$_{trig}$ magnetic moment
  (bottom panel).}
 \label{fig4}
\end{figure}

As far as the valencies are concerned, we find from the LSDA+U+SOC
that both the Co$_{oct}$ and Co$_{trig}$ sites have the 3+
valence. The 4+/2+ scenario (also FMI) is unstable and lies
higher by 84 meV per formula unit (fu). The Co$_{oct}$ ion in the
3+/3+ ground state is definitely LS: as can be seen from Fig.
3(a), the large Co$_{oct}$ $t_{2g}$-$e_{g}$ band gap does not
leave any room for a $3d^{6}$ state other than the one with
$t_{2g}^{6}$, i.e. non-magnetic. The Co$_{trig}$ ion on the other
hand, is in the HS state (S=2). Very interesting is the orbital
occupation for this Co$_{trig}^{3+}$ ion: Fig. 3(b) shows that five
electrons occupy all the spin-up $d_{0}$, $d_{1}$/$d_{-1}$, and
$d_{2}$/$d_{-2}$ orbitals, and that \textit{the sixth electron
resides in the spin-down $d_{2}$ orbital}. 
This is the consequence of the SOC in this
very narrow band system with the tiny crystal field splitting between
the $d_{0}$ and $d_{2}$/$d_{-2}$ levels. 
When the $d_{-2}$ orbital
is forced to be occupied instead, the increasing total-energy
allows us to estimate the SOC strength to be about 70 meV.
As a result, the orbital
contribution to the magnetic moment can be much larger than the 1
$\mu_{B}$ one usually finds for Co or Fe ions in $O_h$
symmetry. An ionic $d_{2}$ state would give an orbital moment of 2 $\mu_{B}$, but here
we find 1.57 $\mu_{B}$ due to covalency and small distortions of
the trigonal prism, see also Table I. Experimentally, Hu
\textit{et al.} have recently found using soft-x-ray magnetic
circular dichroism that Ca$_3$Co$_2$O$_6$ indeed has an orbital
moment which is significantly larger than 1 $\mu_{B}$ \cite{Hu05}.

\begin{table*}[ht]
\caption{Calculated electronic state of Ca$_3$Co$_2$O$_6$; total
energy difference (meV) per formula unit; moments ($\mu_B$) at
each Co$_{oct}$, Co$_{trig}$, and O ion, as well as in the
interstitial region and total magnetic moment per formula unit.
Both Co sites are trivalent in the solutions listed. The high-spin
Co$_{trig}^{3+}$ has either the minority-spin $d_2$ or $d_0$
occupied.} \label{TableI}
\begin{tabular} {l@{\hskip0.5cm}c@{\hskip0.5cm}c@{\hskip0.5cm}c
@{\hskip0.5cm}c@{\hskip0.5cm}c
@{\hskip0.5cm}c@{\hskip0.5cm}c@{\hskip0.5cm}c@{\hskip0.5cm}c
@{\hskip0.5cm}c} \\ \hline\hline
&state&$\Delta$E&Co$_{oct}^{spin}$ &Co$_{oct}^{orb}$ &
Co$_{trig}^{spin}$ &Co$_{trig}^{orb}$ & O  & interstitial &
total \\ \hline LSDA&FMHM&&0.34&&2.70&&0.13&0.13&4.00 \\
\hline
LSDA+U+SOC&FMI($d_2$)&0&0.07&0.09& 2.99&1.57&0.13 &0.13&5.66\\
LSDA+U+SOC&AFI($d_2$)&12&0&0& $\pm$2.99&$\pm$1.57&$\pm$0.13&0&0\\
LSDA+U+SOC&FMI($d_0$)&71&0.002&0.03& 2.96&0.10&0.14&0.14&4.13\\
\hline\hline
\end{tabular}
\end{table*}

Adding the orbital moment to the 4 $\mu_{B}$ from the S=2 spin
contribution, we end up with the total magnetic moment 
of about 5.66 $\mu_{B}$/fu. Experimentally, Maignan \textit{et
al.} \cite{Maignan04} found from magnetization measurements the
saturated total moment close to 5.0 $\mu_{B}$. 

%%THE FOLLOWING DISCUSSION IS TOO EXTENDED, TOO MUCH "LYRICS". IT CAN BE 
%%SHORTENED TO ONE SENTENCE; AND IT WOULD ALSO SAVE SOME SPACE. AND I 
%%REFERRED TO DAI-WANGBO HERE
We may speculate that perhaps a small
part of the sample ($\approx$10\%) is somewhat misaligned, which
due to the anomalously large magnetocrystalline anisotropy as we
will show below, will then not get magnetized. In any case, since
it is easier to find experimental reasons why the sample is not
fully magnetized rather than `over'-magnetized,
it is probably better that the theory slightly overestimates the moment.
It would be especially 
unsatisfactory, if one, for example, would ignore the SOC or 
assume that the sixth electron occupies the $d_{0}$ spin-down
orbital \cite{Dai05}, giving in both cases a total moment of only 4.0
$\mu_{B}$. In fact, as listed in Table I, we find in our
LSDA+U+SOC calculations that the configuration with the sixth
electron forced to occupy the $d_{0}$ spin-down orbital is
unstable by 71 meV/fu as compared to the ground
state. Important is that this analysis is also robust against the
particular choice of $U$ made here. Fig. 4 (bottom panel) shows
that the spin and orbital moments
stay constant within 0.2 $\mu_B$ as long as the system is an
insulator, i.e. when $U$ is varied between 2.5 and 6 eV.

The occupation of the spin-down $d_{2}$ orbital has as a
consequence that the orbital and, due to the SOC, also the spin
contributions to the magnetic moment are oriented along the
$z$-direction, i.e. the $c$-axis. To flip both the spin and
orbital moments into the plane perpendicular to the $c$-axis is
impossible, since this would require that the spin-down
$d_{1}$ and $d_{-1}$ become partially occupied, and we have seen
from the LSDA calculations in Fig. 1(b) that this would cost 
roughly the crystal field energy ($\sim$1 eV). To flip only the spin
moment into the plane but keep the orbital moment along the
$c$-axis is also quite difficult, since this would cost the full
spin-orbit splitting of about 70 meV. Also a third
scenario in which one tries to alter the moment directions by
occupying non-spin-orbit-active spin-down $d_{0}$ orbital is
equally unlikely, since our LSDA+U+SOC calculations show that
this would cost 71 meV, as already mentioned above and in Table
I. All this means that the magnetocrystalline anisotropy is
exceptionally large, and that all the relevant magnetic degrees
of freedom are highly fixed in the $z$-direction, justifying why
the magnetic behavior of Ca$_3$Co$_2$O$_6$ can be well described
by Ising models.

In trying to explain the intra-chain FM exchange
interaction we also have investigated the antiferromagnetic
insulating (AFI) scenario. Here we took the chain as an
alternation of spin-up and spin-down HS-Co$_{trig}$ ions with
LS-Co$_{oct}$ ions in between. Our LSDA+U+SOC
calculations find that the AFI solution, having the same spin and
orbital moments at the HS Co$_{trig}$ ions as the FMI
ground-state, lies above the latter by about 12 meV/fu,
see Table I \cite{note2}. As a result, the intra-chain FM coupling
parameter of each HS Co$_{trig}$ pair can be estimated to be of
the order of 1.5 meV (17 K), assuming a simple Heisenberg model
with S=2. This intra-chain exchange parameter is in reasonable
agreement with the experimentally observed intra-chain Curie
temperature of 24$\pm$2
K~\cite{Aasland97,Kageyama97,Maignan00}.

This result seems at first sight to support the ionic
superexchange model proposed by Fr\'{e}sard \textit{et al.}
\cite{Fresard04}. That model used an \textit{ansatz} for the Co
ions which was confirmed by our LSDA+U+SOC, namely LS
Co$^{3+}_{oct}$ and HS Co$^{3+}_{trig}$ with the $d_{0}$ orbital
half occupied, and found the exchange parameter of about 2 meV,
very close to ours. However, that model considered basically a
set of virtual excitations which are purely ionic in character and
involve essentially only the Co$_{trig}$ and Co$_{oct}$
$3z^{2}-r^{2}$ orbitals, in which case the solution should be AF,
according to the first Goodenough-Kanamori-Anderson rule
\cite{Goodenough63}.

We believe that the intra-chain FM 
interaction is connected with the large contribution of holes
on the oxygens. On the basis of general arguments and also our
calculations, one finds that the actual charge distribution
corresponds rather not to Co$^{3+}$, but to
Co$^{2+}$$\underline{L}$ ($\underline{L}$=oxygen hole), 
although the quantum
numbers are those of Co$^{3+}$. In this case two of such oxygen holes from 
two neighbouring HS-Co$^{3+}_{trig}$ form a triplet state at the 
LS-Co$^{3+}_{oct}$ in between, much the same as two oxygen holes form 
a triplet state around a metal vacancy in ZnO and CaO \cite{Elfimov02}. 
This allows to gain
the full coherent hybridization energy with the empty $e_g$ states of the
Co$^{3+}_{oct}$ ion. An indication for this is the
presence of a covalency-induced spin moment in the
oxygen of about 0.13 $\mu_B$ per ion as well as in the 
LS-Co$^{3+}_{oct}$ of about 0.07 $\mu_B$, see Table I. 
This picture of the origin of the FM intra-chain exchange should be 
confirmed by further studies.

To conclude, we find using LSDA+U+SOC band structure calculations
that Ca$_3$Co$_2$O$_6$ is not a ferromagnetic half-metal but a
Mott insulator with both the octahedral and the trigonal Co ions
being formally trivalent, which settles the valence-state issue. Spin-orbit coupling and unusual
coordination of the trigonal Co ion lead to the occupation of the
$d_{2}$ spin-down orbital, generating a giant orbital
moment (1.57 $\mu_{B}$) and an extremely large magnetocrystalline
anisotropy (70 meV), which explains the Ising character of
the magnetism of this material. The ferromagnetic intra-chain 
interaction is presumably connected with the effect of oxygen holes 
in this small charge-transfer gap system. This also leads to an
appreciable magnetic moment of 0.13 $\mu_{B}$ on each oxygen.

We are grateful to Zhiwei Hu, Antoine Maignan, Vincent Hardy, and
George Sawatzky for stimulating discussions. This research is
supported by the Deutsche Forschungsgemeinschaft through SFB 608.


\begin{thebibliography}{99}
\bibitem{note} Our LSDA+U calculations show that
the insulating ground state with a gap of 0.4 eV possesses 
LS-Co$_{oct}^{3+}$ and HS-Co$_{trig}^{3+}$ having the spin-down
$3z^2-r^2$ ($d_0$) occupied (no orbital moment), which is more
stable by 392 meV/fu than the HM state with the
HS-Co$_{trig}^{3+}$ having the spin-down $x^2-y^2$/$xy$ 
half-filled. Therefore, $U$ determines
the insulating behavior, but the SOC accounts for the Ising 
magnetism, as discussed in the main text.
\bibitem{note2} In the all range of $U$=0---6 eV, the
intrachain FM state is constantly more stable than the AF state,
by 88, 49, 19, 12, 11 meV/fu when $U$=0, 2.5, 4, 5,
6 eV, respectively. For any realistic value of $U$ (about 4---6 eV),
the exchange parameter is estimated to be 1.4---2.4 meV (16---27 K).
\end{thebibliography}
\end{document}